\documentclass[unsortedaddress,showpacs,showkeys]{revtex4}
\usepackage{amsfonts}
\usepackage{amsmath}
\usepackage{amssymb}
\usepackage{bbold} 

\newcommand{\BvK}{Born-von K\'arm\'an}
\newcommand{\Q}{\ensuremath{\mathfrak{Q}}}
\newcommand{\Qc}{\ensuremath{\mathfrak{Q}_c^d}}

\newcommand{\Td}{\ensuremath{{\cal T}_d}}
\newcommand{\TNd}{\ensuremath{{\cal T}_{Nd}}}
\newcommand{\Dd}{\ensuremath{\tilde\Td}}

\newcommand{\Tper}{\ensuremath{{\bf T}_{per}}}
\newcommand{\Dper}{\ensuremath{\tilde{\bf T}_{per}}}

\newcommand{\KBvK}{\ensuremath{K_{\textrm{BvK}}}}
\newcommand{\QBvK}{\ensuremath{\mathfrak{Q}_{\textrm{BvK}}}}
\newcommand{\Qres}{\ensuremath{\mathfrak{Q}_{\textrm{res}}}}

\newcommand{\comment}[1]{}

\begin{document}


\title{Vanishing Integral Relations and Expectation Values for Bloch Functions in Finite Domains}

\author{C. Pacher}
\email{christoph.pacher@arcs.ac.at}
\thanks{phone: +43 1 50550 4165, fax: +43 1 50550 4190}
\author{M. Peev}
\affiliation{ARC Seibersdorf Research GmbH, Donau-City
Stra\ss e 1/4, A-1220 Wien, Austria}
\date{\today}

\begin{abstract}
Integral identities for particular Bloch functions in \textit{finite} periodic systems are derived. All following statements are proven for a finite domain consisting of an integer number of unit cells. It is shown that matrix elements of particular Bloch functions with respect to periodic differential operators vanish identically.
The real valuedness, the time-independence and a summation  property of the expectation values of periodic differential
operators applied to superpositions of specific Bloch functions are derived.
\end{abstract}
 \maketitle

\section{Introduction\label{sec:intro}}



The solutions of periodic systems with periodic boundary conditions (PBC) and of finite periodic systems with complete quantum confinement (CQC), are characterized by a discrete spectrum of the Bloch quasi-momentum $q$. In contrast in \comment{\textit{open}} finite systems with open boundary conditions (OBC), solutions do exist for any real value of $q$, as in an infinite periodic system.

In the literature general orthogonality relations for Bloch waves can be found only for either (i) infinite systems or (ii) systems with either PBC or CQC, but not for systems with OBC. Nevertheless, locally periodic regions of finite length $L$ with open boundary conditions are more realistic models of crystals than systems with periodic boundary conditions which have been studied extensively in the literature. 

In this paper we derive vanishing integral relations valid for particularly chosen Bloch functions occuring in 
\textit{finite} 
periodic systems consisting of an integer number of unit cells with either PBC, CQC or OBC. The latter pertain to finite-size crystal models as well as semiconductor superlattices. These integral relations include orthogonality relations and conditions for the vanishing of matrix elements. The choice of Bloch functions in these relations is directly related to the symmetry group of a system in which periodic boundary conditions are imposed at the ends of the finite periodic potential. However, the relations we derive are not restricted to ``quantized'' values of $q$ which correspond to systems with PBC or CQC only, but hold also for uncountably many well-defined combinations of wave numbers $q$ from the \emph{continuous} quasi momentum space. On this basis we study properties of specific linear combinations of Bloch functions.
Our results indicate that the large body of group theoretical results on systems with PBC can also be applied to more realistic models with no apparent symmetry whatsoever. 

The paper is organized as follows. In Section \ref{sec:Fundamentals} we shortly review some important properties of the solutions of the Schr\"odinger equation in the case of (finite) periodic one-dimensional potentials. In Section \ref{sec:MainResult} we state the fundamental result on vanishing integral relations for finite periodic systems [cf.~Eq.~(\ref{eq:fundamentalEq1}), (\ref{eq:fundamentalEq2}) and (\ref{eq:fundamentalEqGeneral})], which is a generalization of the well known selection rule theorem for PBC. Subsequently in Section \ref{sec:Applications} we discuss applications of this main result to vanishing integrals involving resonant Bloch functions [as defined in Eq.~(\ref{eq:q_res})], vanishing matrix elements of specific Bloch functions and orthogonality relations. Finally in Section \ref{sec:ExpectValues} we consider special properties -- among them time-independence and real valuedness  -- of physically relevant expectation values of Bloch functions and superpositions thereof in open finite systems. Appendix \ref{app:group_theory} contains concepts which are subsequently used for a group theoretical proof of Eq.~(\ref{eq:fundamentalEq1}).

\section{Schr\"odinger equation for (finite) periodic one-dimensional potentials --- Fundamentals \label{sec:Fundamentals}}

It is well known that the fundamental solutions of the time independent one-dimensional Schr\"odinger equation with an infinite periodic potential (with period $d$)
\begin{equation}
\hat H\Psi(x) =\left[-\frac{\hbar^2}{2m}\frac{d^2}{dx^2}+V_{per}(x)\right]\Psi(x)=E\Psi(x), \quad V_{per}(x+d)=V_{per}(x),
\end{equation}
inside an allowed band are Bloch functions \cite{Bloch29} $\Psi^B_{\tilde q}(x)$, given by
\begin{equation}
\Psi^B_{\tilde q}(x)=u_{\tilde q}(x)\exp(i\tilde qx), \quad u_{\tilde q}(x+d)=u_{\tilde q}(x), \quad
{\tilde q}\in\mathbb{R}, \label{eq:PsiBloch_basic}
\end{equation}
where ${\tilde q}$ denotes the energy dependent Bloch wave number or quasi momentum \cite{Kittel87}. 
To obtain the Bloch wave number $q$ in the ``reduced zone scheme'' \cite{Kittel87} the minimal residue of $\tilde{q} \mod 2\pi/d$ is taken. As is well known this ``folds'' the Bloch wave number from the $n$-th to the first Brillouin zone $q\in(-\pi/d,\pi/d]$. 
In this reduced zone scheme the original Brillouin zone number is added as band index $n$. Then Eq. (\ref{eq:PsiBloch_basic}) reads
\begin{equation}
\Psi^B_{\tilde q}(x)=\Psi^B_{n,q}(x)=u_{n,q}(x)\exp(iqx), \quad u_{n,q}(x+d)=u_{n,q}(x), 
\label{eq:PsiBloch}
\end{equation}
which satisfies $\hat H\Psi^B_{n,q}(x)=E_{n,q}\Psi^B_{n,q}(x)$. The relation $E_n(q):=E_{n,q}$ is known as a dispersion relation.
Due to Kramers degeneracy $E_{n,q}=E_{n,-q}$. Therefore any superposition \cite{fn:BandEdgeBloch} 
\begin{equation}
\Psi^B_{n,(q,-q)}(x)=\alpha^+\Psi^B_{n,q}(x)+\alpha^-\Psi^B_{n,-q}(x) \label{eq:BlochSuperpos}
\end{equation} 
with arbitrary $\alpha^+,\alpha^-$ satisfies $\hat H\Psi^B_{n,(q,-q)}(x)=E_{n,q}\Psi^B_{n,(q,-q)}(x)$. 

\subsection{Periodic boundary conditions}
\label{sec:PBC}
Since the introduction of Bloch functions their properties have been studied in detail mostly for systems with periodic (\BvK = BvK) boundary conditions, i.e. $\Psi(x)=\Psi(x+Nd)$. The BvK boundary conditions imply a quantization of the Bloch wave number in Eq.~(\ref{eq:PsiBloch}) in the form of
\begin{equation}
q_\textrm{BvK}^{(k)}=2k\pi/Nd, 
\quad k\in K_\textrm{BvK}:=
\left\{-\lfloor (N-1)/2 \rfloor,\dots,-1,0,1,\dots,\lfloor N/2\rfloor\right\} \subset\mathbb{Z},
\label{eq:q_BvK}
\end{equation}
but no restriction on $\alpha^+$ and $\alpha^-$ in Eq.~(\ref{eq:BlochSuperpos}). 
The set 
\begin{equation}
\QBvK:=\left\{q^{(k)}_{\textrm{BvK}}, 
k\in \KBvK\right\},
\quad |\QBvK|=N \label{eq:setQ}
\end{equation}
forms a finite cyclic group $\langle \QBvK,\oplus \rangle$, where the group operation $\oplus$ is addition modulo $2\pi/d$ with minimal residue taken, i.e.
\begin{equation}
q_1\oplus q_2 := \left\lbrace
\begin{array}
{cl}%
q_1+q_2 + 2\pi/d &\quad \textrm{if}\quad q_1+q_2\leq -\pi/d, \\
q_1+q_2 & \quad \textrm{if}\quad  -\pi/d <q_1+q_2 \leq \pi/d, \\
q_1+q_2 - 2\pi/d &\quad \textrm{if}\quad q_1+q_2> \pi/d. \label{eq:qoplus}
\end{array}
\right.
\end{equation} 

\subsection{Complete quantum confinement in finite periodic systems} \label{sec:CQC}
Recently, interest in studying properties of \textit{finite} locally periodic quantum systems has turned up \cite{Sprung93,Sprung00,Ren01,Pereyra02,Ren02,Pacher03,Pacher05,Pereyra05}. One option to address such finite systems is to introduce complete quantum confinement (CQC)\cite{Ren01,Pereyra05}, i.e. $\Psi(0)=\Psi(Nd)=0$. 
The wave number quantization of the real Bloch wave number in Eq.~(\ref{eq:BlochSuperpos}) in this case yields a set
\begin{eqnarray}
\mathfrak{Q}_\textrm{CQC} &:=& \left\{q_\textrm{CQC}^{(k)}= k\pi/Nd, 
\quad k\in \{-N+1,\dots,-2,-1,1,2,\dots,N-1\} \right\}, \label{eq:q_cqc} \\
\quad |\mathfrak{Q}_\textrm{CQC}|&=&2N-2, \nonumber 
\end{eqnarray}
but with the additional restriction: 
$\alpha^+ + \alpha^-= 0$, if $u_{n,q}(0)$ is choosen to be real \cite{fn:SurfaceStates}. Note that due to this constraint the wave numbers $q_\textrm{CQC}^{(k)}$ and $q_\textrm{CQC}^{(-k)}$ always jointly appear in Eq.~(\ref{eq:BlochSuperpos}). The set $\mathfrak{Q}_\textrm{CQC}$ coincides with the set $\Qres$ below [cf.~Eq.(\ref{eq:q_res})] and can analogously be closed to a group.

\subsection{Open boundary conditions for finite periodic systems}
\label{sec:OBC}
The potential of an \textit{open}, finite $N$-fold periodic systems \cite{Sprung93,Pereyra02,Pacher03,Pacher05} is
\begin{equation}
V(x)=\left\lbrace
\begin{array}
{cl}%
0 & \textrm{if } x\leq0,\\
V_{per}(x) &\textrm{if } 0\leq x\leq L=Nd,\\
0 & \textrm{if } x\geq L.
\end{array}
\right. \label{eq:PotOpenSysten}
\end{equation}
The (rigged) Hilbert space of this problem is not restricted to the interval $[0,L]$ but extended to the whole real line. 
In an open system, solutions (also called scattering states) in the form of Eq.~(\ref{eq:BlochSuperpos}), exist for $0\le x \le L$ for all Bloch wave numbers in $(-\pi/d,\pi/d]$:
\begin{equation}
V(x)=\left\lbrace
\begin{array}
{cl}%
A_{L}\exp[ik(E_{n,q})x]+B_{L}\exp[-ik(E_{n,q})x] & \textrm{if } x\leq0,\\
\alpha^+\Psi^B_{n,q}(x)+\alpha^-\Psi^B_{n,-q}(x) &\textrm{if } 0\leq x\leq L=Nd,\\
A_{R}\exp[ik(E_{n,q})x]+B_{R}\exp[-ik(E_{n,q})x] & \textrm{if } x\geq L.
\end{array}
\right. \label{eq:WaveOpenSysten}
\end{equation}
Hence the Bloch wave number $q$ is not quantized, and there are no restrictions on the coefficients $\alpha^\pm$, as in the PBC system.

By considering resonant states (with a transmission probability of unity and a reflection probability of zero) of an open system with a potential given by Eq.~(\ref{eq:PotOpenSysten}) a new kind of ``quantization'' arises. 
The set of discrete $q$-values that belong to Fabry-P\'{e}rot resonances \cite{Pacher01} of the open system is given by \cite{Sprung93,Pacher01,Pereyra02,Pacher03}
\begin{eqnarray}
\Qres &:=&\left\{q^{(k)}_{res}= k\pi/Nd=2k\pi/2Nd, 
\quad k\in
\{-N+1,\dots,-2,-1,1,2,\dots,N-1\}\right\}, \label{eq:q_res} \\
|\Qres|&=&2N-2. \nonumber
\end{eqnarray}
This quantization corresponds to a BvK system with $2N$ periods. The ``doubling'' of the number of periods is fundamentally related to the fact that resonant wave functions fulfill the relations \cite{Sprung93,Pacher03} $\Psi(0)=\pm\Psi(Nd)$ which means that $\Psi(0)=\Psi(2Nd)$. It is possible that additional non-Fabry-P\'{e}rot resonances exist which are due to transmission resonances in the \emph{unit cell} \cite{Sprung93,Pacher01,Pacher03,Pacher05} but these are outside the scope of the current paper.


Note that the union of \Qres\ with the set of $q$ values belonging to the band edges, i.e. $\tilde\Q{}_\textrm{res}:=\Qres \cup \{0,\pi/d\}$, forms together with the operation $\oplus$ a group and $|\tilde\Q{}_\textrm{res}|=2N$. Obviously, as $2q_\textrm{res}^{(k)}=q_\textrm{BvK}^{(k)}$, $\tilde\Q{}_\textrm{res}=\QBvK \cup q_\textrm{res}^{(1)}\oplus\QBvK$ 

\section{Vanishing integrals involving products of Bloch functions\label{sec:MainResult}}
In this paragraph we state fundamental integral identities which emerge in the analysis of Bloch waves in finite periodic systems (see e.g. Ref.~\cite{Pacher05}). 
To our best knowledge these identities are presented below in a general form for the first time.

The group $\langle \QBvK,\oplus \rangle$ [cf. Eq.~(\ref{eq:setQ})] is a discrete (normal) subgroup of the compact continuous one-dimensional rotation group $\langle \Qc,\oplus \rangle$, where
\begin{equation}
\Qc:=(-\pi/d,\pi/d]. \label{eq:Qc}
\end{equation}
The following fundamental integral identity holds:
\begin{equation}
\bigoplus_{r=1}^R q_r\in \QBvK \backslash \{0\} \Rightarrow\int_0^{Nd} dx \frac{d^j f_{per}(x)}{dx^j} \prod_{r=1}^R \frac{d^{j_r}  \Psi^B_{{n_r},q_{r}}(x)}{dx^{j_r}}=0, 
\label{eq:fundamentalEq1}
\end{equation}
where $q_r\in \Qc$, $\bigoplus$ denotes a sum with respect to the group operation in \Qc\ and $f_{per}(x+d)=f_{per}(x)$ denotes an arbitrary $d$-periodic function.

Equation~(\ref{eq:fundamentalEq1}) can trivially be extended to
\begin{equation}
\left(\bigoplus_{r=1}^{R}q_r\right)\oplus
\left(\bigoplus_{s=1}^{S}(-q_s)\right)\in  \QBvK \backslash \{0\} \Rightarrow
\int_0^{Nd} dx \frac{d^j f_{per}(x)}{dx^j} 
\prod_{r=1}^{R} \frac{d^{j_r} \Psi^B_{{n_r},q_{r}}(x)}{dx^{j_r}}
\prod_{s=1}^{S} \frac{d^{j_s}  [\Psi^B_{{n_s},q_{s}}(x)]}{dx^{j_s}}^*=0, 
\label{eq:fundamentalEq2}
\end{equation} 
where we make use of the fact that $\Psi^{B*}_{n,q}$ and $\Psi^B_{n,-q}$ differ at most by a constant phase factor.

Most generally one can extend Eq.~(\ref{eq:fundamentalEq1}) to  \begin{equation}
\int_0^{Nd} dx I(x)=0, \label{eq:fundamentalEqGeneral}
\end{equation}
where the integrand $I(x)$ is any sequence of terms belonging to the set 
\begin{equation}
\{d/dx,\Psi^B_{n_r,q_r}(x),[\Psi^B_{n_s,q_s}(x)]^*,f^{per}_t(x)\},
\end{equation}
where additionally pairs of parentheses can be inserted appropriately and operators act from right to left as usual. Once again 
\begin{equation}
\left(\bigoplus_{r} q_r\right) \oplus \left(\bigoplus_{s} (-q_s)\right)\in  \QBvK \backslash \{0\}.  \label{eq:qconditionlast}
\end{equation} 
For example valid integrands are $d/dx \Psi^B_{n_1,q_1}(x) f_1^{per}(x)\Psi^B_{n_2,q_2}(x) [d/dx \Psi^B_{n_3,q_3}(x)] f_2^{per}(x)$ (with $q_1\oplus q_2\oplus q_3\in \QBvK \backslash \{0\}$), where the first $d/dx$ acts on the rest of the term, or $[\Psi^B_{n_1,q_1}(x)]^* d^2/dx^2 f_1^{per}(x) \Psi^B_{n_2,q_2}(x)$  (with $-q_1\oplus q_2\in \QBvK \backslash \{0\}$) where $d^2/dx^2$ acts on the product $f_1^{per}(x) \Psi^B_{n_2,q_2}(x)$.
This last generalization follows from Eq.~(\ref{eq:fundamentalEq2}) by applying the product rule of differentiation. 

The proof for Eq.~(\ref{eq:fundamentalEq1}) can be seen from two perspectives. 
On the one hand, it is well known \cite{Kittel87} that 
\begin{equation}
\int_0^L dx f_\textrm{per}(x)\exp\left(iq_\textrm{BvK}^{(k)}x\right) = 0, \quad \textrm{if } k\ne0,
\end{equation} 
and obviously
\begin{equation}
\int_0^L dx f_\textrm{per}(x)\exp\left(i\sum_r q_rx\right) = 0, \quad \textrm{if }\sum_r q_r\in \QBvK \ne0, \label{eq:Kittel2}
\end{equation} 
for otherwise arbitrary wave numbers $q_r$ from \Qc. Equations~(\ref{eq:fundamentalEq1}) and (\ref{eq:Kittel2}) are equivalent taking into account the translational invariance of the differential operator. Normally  relations such as Eq.~(\ref{eq:Kittel2}) are widely used in solid state physics but all summands in $\sum_r q_r$ are restricted to belong to \QBvK\ as only PBCs are considered. 
Eq.~(\ref{eq:fundamentalEq1}) extends the well known selection rules theorem to uncountably many well-defined combinations of wave numbers $q$ from the domain of \emph{continuous} quasi momenta for finite locally periodic open systems.

In Appendix A we discuss the proof of Eqs.~(\ref{eq:fundamentalEq1}-\ref{eq:fundamentalEqGeneral}) from a rigorous group theoretic point of view.

\section{Applications to specific cases}
\label{sec:Applications}

\subsection{Vanishing integrals involving two resonant Bloch functions}
\label{sec:VanishingInt2ResBloch}
Identities involving two resonant Bloch functions $\Psi^B_{n,q^{(k)}_{res}}$ and $\Psi^B_{n',q_{res}^{(k')}}$ which follow from Eq.~(\ref{eq:fundamentalEq1}) 
are
\begin{equation}
q_\textrm{res}^{(k)} \oplus q_\textrm{res}^{(k')} \in\QBvK \setminus \{0\} \Rightarrow \int_0^{Nd} dx \frac{d^j f_{per}(x)}{dx^j} \frac{d^l  \Psi^B_{n,q^{(k)}_{res}}(x)}{dx^l} \frac{d^m  \Psi^B_{n',q_{res}^{(k')}}(x)}{dx^m}=0,
\label{eq:per_DPsiDPsi}
\end{equation}
where $j,l,m \in \mathbb{N} \cup \{0\}$.

For example an extension of Eq.~(A2) in Ref.~\cite{Pacher05} to position dependent effective masses $m(x)$ is of the form of Eq.~(\ref{eq:per_DPsiDPsi}) with $k=k', j=0, f_{per}(x)=1/m(x), l=0, m=1$.


\subsection{Vanishing matrix elements for specific Bloch functions}
\label{sec:VanishingMatrixElSpecBloch}

Equation~(\ref{eq:fundamentalEq2}) allows to consider matrix elements of the form
$\langle \Psi^B_{n_i,q_i}\vert \hat A \vert \Psi^B_{n_j,q_j} \rangle_{Nd}$, where $\hat A$ is a periodic differential operator with respect to space, i.e.
\begin{equation}
\hat A(x)=\sum_l a_l(x)\frac{d^l}{dx^l}, \qquad a_l(x+d)=a_l(x) \label{eq:diffOp}
\end{equation}
and $\langle \cdot \rangle_{Nd}$ denotes projection onto the finite space region $[0,Nd]$, i.e.
\begin{equation}
\langle \Psi\vert \hat A \vert \Phi \rangle_{Nd}:=\langle \Psi\vert \hat A {\hat P}_{\text{FPS}} \vert \Phi \rangle,
\end{equation}
where 
${\hat P}_{\text{FPS}}=\int_{0}^{Nd}dx\,|x\rangle\langle x|$.
Since the rotation group \Qc\ [cf. Eq.~(\ref{eq:Qc})] can be trivially decomposed in cosets with respect to \QBvK:
\begin{equation}
\Qc=\bigcup_{q_c^{Nd}} \left( q_c^{Nd} \oplus \QBvK \right), \quad q_c^{Nd} \in (-\pi/Nd,\pi/Nd]=(-q_\textrm{res}^{(1)},q_\textrm{res}^{(1)}],
\end{equation}
the following relation holds for any operator $\hat A$ of the form given in Eq.~(\ref{eq:diffOp}):  All matrix elements of $\hat A$ for two different Bloch waves (of the same or of different bands) with wave numbers $q_i$ and $q_j$ in the same coset of \QBvK\ vanish, i.e.
\begin{equation}
q_i\ne q_j; q_i,q_j\in q_c^{Nd}\oplus\QBvK \Rightarrow \langle \Psi^B_{n_i,q_i}\vert \hat A \vert \Psi^B_{n_j,q_j} \rangle_{Nd}=0. \label{eq:interference}
\end{equation} 
Of course, the condition is equivalent to $q_j-q_i=\frac{2k\pi}{Nd}\ne0$.


A special case of this situation is the vanishing of matrix elements of (Kramers-) degenerated resonant solutions which emerge in the fundamental solution of the open or bounded finite periodic system,
\begin{equation}
\langle \Psi^B_{n, q_{res}^{(\pm k)}}\vert \hat A \vert \Psi^B_{n, q_{res}^{(\mp k)}} \rangle_{Nd} =0, \label{eq:matElementqres}
\end{equation}
since $q_{res}^{(-k)},q^{(k)}_{res}\in q^{(k)}_{res}\oplus\QBvK$.

\subsection{Orthogonality relations}
\label{sec:OrthoRels}

Setting $\hat A=\mathbb{1}$ in Eq.~(\ref{eq:interference}) we obtain the following orthogonality relation: Two different Bloch waves (of the same or of different bands) with wave numbers $q_i$ and $q_j$ in the same coset of \QBvK\ are orthogonal, i.e.
\begin{equation}
q_i\ne q_j; q_i,q_j\in q_c^{Nd}\oplus\QBvK \Rightarrow \langle \Psi^B_{n_i,q_i} \vert \Psi^B_{n_j,q_j} \rangle_{Nd}=0. \label{eq:orthogonal1}
\end{equation} 
Again, the condition is equivalent to $q_j-q_i=\frac{2k\pi}{Nd}\ne0$.

Similarly to Eq.~(\ref{eq:matElementqres}) we get orthogonality of (Kramers-) degenerated resonant solutions, i.e.
\begin{equation}
\langle \Psi^B_{n, q_{res}^{(\pm k)}} \vert \Psi^B_{n, q_{res}^{(\mp k)}} \rangle_{Nd} =0. \label{eq:orthogonalqres}
\end{equation}

\section{Expectation values for specific superpositions of Bloch functions}
\label{sec:ExpectValues}


We consider (normalized) superpositions composed of Bloch waves from arbitrary bands (with band indices $n_i$) for which all reduced Bloch wave numbers $q_i$ belong to the same coset of $\QBvK$ but are different, i.e.
\begin{eqnarray}
|\Psi\rangle  = \sum_i \alpha_{i}|\Psi^B_{n_i,q_i}\rangle, \quad
\sum_{i} |\alpha_{i}|^2=1, \text{ where } (\forall i: q_i\in q_c^{Nd}\oplus\QBvK) \wedge (i\ne i' \Rightarrow q_i\ne q_{i'}). \label{Eq:SuperpositionQ}
\end{eqnarray}
These superpositions can consist at most of $N$ Bloch waves with corresponding $N$ different $q$-values $q_1,\dots,q_{N}\in q_c^{Nd}\oplus\QBvK$. The number of different bands involved in this superposition can be any number between 1 and $N$.

Superpositions of the form given in Eq.~(\ref{Eq:SuperpositionQ}) have a special property following directly from Eq.~(\ref{eq:interference}): The expectation value of a periodic differential operator restricted to a finite domain of $N$ periods [cf.~Eq.~(\ref{eq:diffOp})] is equal to the sum of the weighted expectation values of that operator with respect to the individual Bloch waves constituting the superposition:
\begin{equation}
\langle \hat A \rangle_{Nd}=\Big\langle \sum_{i}\alpha_{i}\Psi^B_{n_i,q_i}\Big\vert \hat A \Big\vert  \sum_{j}\alpha_{j}\Psi^B_{n_j,q_j} \Big\rangle_{Nd} =
\sum_{i}\langle\alpha_{i}\Psi^B_{n_i,q_i}\vert \hat A \vert \alpha_{i}\Psi^B_{n_i,q_i} \rangle_{Nd} =
\sum_{i} |\alpha_{i}|^2 \langle \Psi^B_{n_i,q_i}\vert \hat A \vert \Psi^B_{n_i,q_i} \rangle_{Nd}.
\label{eq:expectidentity}
\end{equation}

Note that these expectation values are time independent. Indeed the time dependence of the wave function $|\Psi(t)\rangle  = \sum_i \alpha_i(0) \exp(-\iota E(n_i,q_i)t/\hbar)|\Psi^B_{n_i,q_i}\rangle$ 
is not reflected in the expectation value:
\begin{equation}
\langle \hat A(t)\rangle_{Nd}=
\sum_{i} |\alpha_{i}(0)|^2 \langle \Psi^B_{n_i,q_i}\vert \hat A \vert \Psi^B_{n_i,q_i} \rangle_{Nd}=\langle \hat A(0)\rangle_{Nd}.
\label{eq:expectidentitytime}
\end{equation}
\subsection{A set of operators which have real expectation values with respect to Bloch functions in $\mathcal{L}^2([0,Nd])$}
\label{sec:OpsWithRealExpVals}
\newcommand{\opA}{\hat D_H^{(n)}}
\newcommand{\opB}{\hat B}

Next we concentrate on a set of operators $\hat A_H$ that are
known to be self-adjoint with domains that are dense in
$\mathcal{L}^2(\mathbb{R})$. First we note that such operators are
in general non-self-adjoint over $\mathcal{L}^2([0,Nd])$.
As an example consider the momentum operator $\hat A_H=\hat p=-i\hbar
\frac{d}{dx}$ on the domain $[0,Nd]$. Partial integration gives
\begin{eqnarray}
\langle \Psi_{n_i,q_i}^B |\hat p|\Psi_{n_j,q_j}^B \rangle_{Nd}& = &-i\hbar \int_0^{Nd} [\Psi_{n_i,q_i}^B(x)]^* [\Psi_{n_j,q_j}^{B}(x)]^\prime dx\nonumber \\
&=& -i\hbar \left[(\Psi_{n_i,q_i}^{B*} \Psi_{n_j,q_j}^B)(Nd)-(\Psi_{n_i,q_i}^{B*} \Psi_{n_j,q_j}^B)(0)\right]
+ \langle \Psi_{n_j,q_j}^B |\hat p|\Psi_{n_i,q_i}^B \rangle^*_{Nd}. \label{eq:nonHermit}
\end{eqnarray}
The operator is clearly non-symmetric (and therefore
non-self-adjoint \cite{Reed_Simon}) as the boundary term on the
right hand side does not vanish for arbitrary Bloch waves.
Therefore such operators do not have real expectation values with respect to arbitrary linear combinations of Bloch functions. 

The set of 
 operators
which we study is built by (a) operators of the form
$\opA=\left(-i\frac{d}{dx}\right)^n, n\in\mathbb{N}$, (b) real
valued periodic functions $\opB: \mathbb{R}\rightarrow\mathbb{R},
\opB(x)=\opB(x+d)$, (c) symmetrized products of operators $\opA$
and $\opB$, e.g. $\hat A_H=\opA \opB \opA$, $\hat A_H=\opB \opA
\opB$, $\hat A_H=\frac{1}{2}\left(\opA \opB + \opB \opA\right)$
and similar products with more operators. Such operators appear frequently in the context of periodic semiconductor heterostructures (superlattices). For a periodic potential $V(x)$ and a periodic effective mass $m(x)$, the momentum operator $\hat p$, the
velocity operator $\hat v$, and the most commonly used Hamiltonian
$\hat H$ are given by $\hat p=\hbar \hat D_H^{(1)}$, $\hat
v=\frac{\hbar}{2}\left(\hat D_H^{(1)}m^{-1}(x)+m^{-1}(x)\hat
D_H^{(1)}\right)$, $\hat
H=\frac{\hbar^2}{2}D_H^{(1)}m^{-1}(x)D_H^{(1)}+V(x)$,
respectively.

We now prove that the expectation values of $\hat A_H$ with
respect to Bloch functions  in $\mathcal{L}^2([0,Nd])$, i.e. $\langle \Psi^B_{n,q}|\hat A_H|\Psi^B_{n,q}\rangle_{Nd}$, are still real numbers.
To this end we start with the first order differential operator
$\hat D_H^{(1)}=-i \frac{d}{dx}$. 
Inserting $\langle x|\Psi^B_{n_i,q_i}\rangle=\langle x|\Psi^B_{n_j,q_j}\rangle=\langle x|\Psi^B_{n,q}\rangle=u_{n,q}(x)\rangle\exp(i q x)$ into 
Eq.~(\ref{eq:nonHermit})
leads to a purely periodic function $u_{n,q}^*u_{n,q}(x)$ in the boundary term, which therefore vanishes. Thus 
\begin{equation}
\langle \Psi_{n,q}^B |\hat D_H^{(1)}|\Psi_{n,q}^B \rangle_{Nd}=\langle \Psi_{n,q}^B |\hat D_H^{(1)}|\Psi_{n,q}^B \rangle_{Nd}^*\Leftrightarrow \langle \Psi^B_{n,q}|\hat D_H^{(1)}|\Psi^B_{n,q}\rangle_{Nd}\in\mathbb{R}.
\end{equation} 

For an $n$-th order differential operator $\hat A_H$ iterative partial
integration leads to $n$ boundary terms. Regardless of the order $n$
the exponential terms $\exp(\mp iqx)$ cancel and purely
periodic functions remain again. Thus 
\begin{equation}
\langle \Psi^B_{n,q}|\hat A_H|\Psi^B_{n,q}\rangle_{Nd}\in\mathbb{R}.
\end{equation} 
As a consequence, together with Eq.~(\ref{eq:expectidentity}), the operators which we studied in this section have real expectation values for superpositions of the form of Eq.~(\ref{Eq:SuperpositionQ}), i.e. 
\begin{eqnarray}
(\forall i: q_i\in q_c^{Nd}\oplus\QBvK) \wedge (i\ne i' \Rightarrow q_i\ne q_{i'}) \Longrightarrow \nonumber \\
\Big\langle \sum_{i}\alpha_{i}\Psi^B_{n_i,q_i}\Big\vert \hat A_H \Big\vert  \sum_{j}\alpha_{j}\Psi^B_{n_j,q_j} \Big\rangle_{Nd} =
\sum_{i} |\alpha_{i}|^2 \langle \Psi^B_{n_i,q_i}\vert \hat A_H \vert \Psi^B_{n_i,q_i} \rangle_{Nd}\in \mathbb{R}.
\end{eqnarray}


\subsection{Resonant superpositions}
\label{sec:ResSuperpos}

Choosing one of the $q_i$'s in the superposition Eq.~(\ref{Eq:SuperpositionQ}) to be a resonant Bloch wave number $q^{(k)}_{res}$, results in a linear combinations of
resonant states composed of either even or odd $k$ values only:
\begin{equation} \label{eq:superposManyRes}
|\Psi\rangle=\sum_{k\in\mathfrak{K}}\alpha_{k}|\Psi^B_{n_k,q^{(k)}_{res}}\rangle,
 \quad \sum_{k} |\alpha_{k}|^2=1,
\end{equation}
where $\mathfrak{K}$ depends on whether $N$ is even or odd: For $N=2m$ we have the set of possible even $k$ values $\mathfrak{K}=\{k\in 2\mathbb{Z},\; -N+2\le k\le N\}$ and the set of possible odd $k$ values $\mathfrak{K}=\{k\in 2\mathbb{Z}\negthinspace+\negthinspace1 ,\; -N+1\le k\le N-1\}$. For $N=2m+1$ we have $\mathfrak{K}=\{k\in 2\mathbb{Z},\; -N+1\le k\le N-1\}$ and $\mathfrak{K}=\{k\in 2\mathbb{Z}\negthinspace+\negthinspace1 ,\; -N+2\le k\le N\}$, resp. 
 

Then
\begin{equation}
\langle \hat A\rangle_{Nd}=\sum_{k\in\mathfrak{K}}|\alpha_{k}|^2 \langle\Psi^B_{n_k,q^{(k)}_{res}}|\hat A|\Psi^B_{n_k,q^{(k)}_{res}}\rangle_{Nd}\in \mathbb{R}. \label{eq:expSuperposManyRes}
\end{equation}
Note that we have implicitely included the band edge functions $|\Psi^B_{n_k,q^{(0)}_{res}}\rangle$, $|\Psi^B_{n_k,q^{(N)}_{res}}\rangle$ in the decomposition which are strictly speaking non-resonant, but their corresponding Bloch wave numbers belong to the extended group $\tilde{\mathfrak{Q}}_\textrm{res}$.

For a CQC system we have seen that (cf. Section \ref{sec:CQC}) $q_\textrm{CQC}^{(k)}=q_\textrm{res}^{(k)}$ and that  $|\alpha_k|=|\alpha_{-k}|$. Consequently  
\begin{equation}
\langle \hat A\rangle_{Nd}=
\sum_{\substack{1\le k \le N-1 \\ k\in\mathfrak{K}}}
|\alpha_{k}|^2 
\left(\langle\Psi^B_{n_k,q^{(k)}_{CQC}}|\hat A|\Psi^B_{n_k,q^{(k)}_{CQC}}\rangle_{Nd}+
\langle\Psi^B_{n_{k},q^{(-k)}_{CQC}}|\hat A|\Psi^B_{n_{k},q^{(-k)}_{CQC}}\rangle_{Nd}\right)
\in \mathbb{R}, \label{eq:expSuperposCQC}
\end{equation}
and operators $\hat A$ that fulfill 
\begin{equation}
\langle\Psi^B_{n_k,q^{(k)}_{CQC}}|\hat A|\Psi^B_{n_k,q^{(k)}_{CQC}}\rangle_{Nd}=-
\langle\Psi^B_{n_{k},q^{(-k)}_{CQC}}|\hat A|\Psi^B_{n_{k},q^{(-k)}_{CQC}}\rangle_{Nd}                   \end{equation} 
have necessarily a vanishing expectation value $\langle \hat A\rangle_{Nd}$.

\subsection{Velocity Expectation Value at Resonance}
\label{sec:VeloExpValue}
Equation~(\ref{eq:expSuperposManyRes}) shows that the velocity expectation value for a linear combination of several resonant Bloch waves, Eq.~(\ref{eq:superposManyRes}), is given by

\begin{equation} \label{eq:expResVelocity}
\langle \hat v \rangle_{Nd}=\sum_{k\in\mathfrak{K}}|\alpha_{k}|^2v_g(n_k,q^{(k)}_{res})\in \mathbb{R},
\end{equation}
where $v_g(n,k)=\hbar ^{-1}\partial E_{n,q}/\partial q$ denotes the group velocity, which is the velocity expectation value of a Bloch wave.



For a superposition of  two energetically degenerated resonant solutions only, i.e.
\begin{equation}
|\Psi\rangle=\alpha_+|\Psi^B_{n,q^{(k)}_{res}}\rangle+\alpha_-|\Psi^B_{n,-q^{(k)}_{res}}\rangle, \quad |\alpha_+|^2 + |\alpha_-|^2 =1, \label{eq:superposRes}
\end{equation}
Eq.~(\ref{eq:expResVelocity}) and $v_g(n,-q)=-v_g(n,q)$ show that the velocity expectation value is simply given by
\begin{equation}
\langle \hat v \rangle_{Nd}=\left(|\alpha_+|^2-|\alpha_-|^2\right)v_g(n,q^{(k)}_{res})\in \mathbb{R}.
\end{equation}
This result has been obtained in Ref.~\cite{Pacher05} by using rather elaborate calculations.
For a CQC systems, where $|\alpha_+|=|\alpha_-|$, $\langle \hat v \rangle_{Nd}=0$ is obviously fulfilled as it should be.




\section{Conclusions}
\label{sec:Conclusions}

In this paper we have discussed different properties of Bloch waves in finite periodic systems with an integer number of unit cells related to their Bloch wave numbers. 
We have summarized  wave number quantizations and their mutual interrelations for finite periodic systems 
in three different cases: periodic boundary conditions [PBC, Eq.~(\ref{eq:q_BvK})], complete quantum confinement [CQC, Eq.~(\ref{eq:q_cqc})] and resonant states in open systems  [OBC, Eq.~(\ref{eq:q_res})]. Further, useful vanishing integral identities [cf. Eqs.~(\ref{eq:fundamentalEq1})-(\ref{eq:qconditionlast})] for open finite periodic systems, which are generalizations of well known similar identities for PBC systems have been derived. The main innovation in this respect is related to the fact that symmetry properties of the PBC system manifested by the special role of the set \QBvK\ [cf. Eq.~(\ref{eq:setQ})] reappear in finite systems despite their lack of obvious symmetry. 
We have shown that all matrix elements of periodic differential operators [as defined in Eq.~(\ref{eq:diffOp})] -- restricted to a finite domain with an integer number of unit cells -- for two different Bloch waves (of the same or of different bands) with wave numbers in the same coset of \QBvK\ vanish. Consequently these two Bloch waves must also be orthogonal. Finally we have considered superpositions of Bloch waves with wave numbers that belong to the same coset of \QBvK. The expectation value of the discussed  periodic differential operators [cf.~Eq.~(\ref{eq:diffOp})] with respect to these superpositions is time independent, real-valued and equal to the sum of the weighted expectation values with respect to the individual Bloch waves constituting the superposition. In addition some specific applications of these general results are presented.


\begin{acknowledgments}
It is a great pleasure to thank R.~Dirl and P.~Kasperkovitz for reading a previous version of the manuscript and for interesting and fruitful discussions and D.~W.~Sprung for reading and commenting the manuscript.
\end{acknowledgments}

\begin{appendix}

\section{Group theoretical proofs} 
\label{app:group_theory}
We start by introducing some appropriate general group theoretical concepts and then use the symmetry properties of the problem at hand for performing the proof. 

\subsection{Relevant groups and irreducible representations thereof}
\label{app:RelGroupsAndIrreps}
The symmetry group of (one-dimensional) infinite periodic systems with period $d$ is the one-dimensional discrete translation group \Td, which is generated by an elementary translation $d$:
\begin{equation}
\Td=d\ \mathbb Z, \quad n,m\in \mathbb Z, g_n=nd,g_m=md \in \Td\Rightarrow g_n*g_m=(n+m)d=g_{n+m}.
\end{equation} 
As \Td\ is abelian it is well known that its irreducible representations are one-dimensional and are of the form
\begin{equation}
D_{\Td}^q(g_n)=\exp(iqnd), \quad q\in (-\pi/d,\pi/d]. \label{Eq:IrrepsTd}
\end{equation} 
The irreducible representations of \Td\ form an abelian group \Dd\ with the tensor product as group operation:
\begin{eqnarray}
D_{\Td}^{q_1}(g_n)\bigotimes D_{\Td}^{q_2}(g_n) &=& \exp(iq_1nd)\exp(iq_2nd)= 
\exp[i(q_1+q_2)nd]=\exp[i(q_1\oplus q_2)nd]= \nonumber \\
&=& D_{\Td}^{q_1\oplus q_2}(g_n),
\end{eqnarray} 
where $q_1 \oplus q_2$ denotes the minimal residue of $(q_1+q_2) \!\!\mod 2\pi/d$, or in physical terms, the sum $q_1 + q_2$ folded back into the first Brillouin zone $(-\pi/d,\pi/d]$ [cf. Eq.~(\ref{eq:qoplus})].

Indeed, (i) the tensor product of two different irreducible representations of this group is also irreducible, (ii) this tensor product is associative and commutative, (iii) the unit element is given by the trivial representation $D_{\Td}^{0}=1$, whereas (iv) the inverse element of $D_{\Td}^{q}$ is its complex conjugate $D_{\Td}^{-q}$.

Clearly the group \Dd\ and \Qc\ [cf.~Eq.~(\ref{eq:Qc})] are isomorphic through the mapping $q\leftrightarrow \exp(iq\ \circ)$.

We now introduce the symmetry group of a system having translation symmetry \Td\ in which additionally periodic BvK boundary conditions are imposed. This symmetry group is the cyclic factor group $\Tper=\Td/\TNd$, where $\TNd$ is a subgroup of \Td\ generated by the elementary translation $Nd$.

$\Tper$ is explicitly defined as
\begin{eqnarray}
\Tper & = & \{ h_0=0, h_1=d, h_2=2d, \ldots,
h_{N-1}=(N-1)d\}, \nonumber \\
h_i \boxplus h_j & = & (i+j)d \!\!\mod Nd, \label{eq:T_d}
\end{eqnarray}
where $\boxplus$ is the group operation in $\Tper$. All translations of the form $mNd$, $m\in \mathbb{Z}$ act trivially within $\Tper$. 

The irreducible representations of the cyclic group $\Tper$ are given by
\begin{equation}
D_{\Tper}^{\bar{q}_k}(h_l)=\exp(i\bar{q}_kld), \quad \bar{q}_k=\frac{2k\pi}{Nd},\quad 0\le k,l \le N-1. \label{Eq:IrrepsTper}
\end{equation} 
Analogous to the case of the group \Td\ the irreducible representations of \Tper\ form an abelian group \Dper\ which is isomorphic to \QBvK. As a side note we mention that -- since \Tper\ is finite and abelian -- it is also isomorphic to \Dper. 

The irreducible representations of $\Td$ with $q=q^{(k)}_{\textrm{BvK}}$ are irreducible representations of $\Tper$. This follows from the fact that (i) the set of all possible indices $\bar{q}_k$ of irreducible representations of \Tper\ coincides with the set $\QBvK$ of all $q^{(k)}_{\textrm{BvK}}$, cf. Eq.~(\ref{eq:setQ}), and (ii) any general translation $g_n=nd\in \Td$ can be represented as $nd=ld+mNd$, where $ld\in \Tper$. 

Indeed 
\begin{equation}
D_{\Td}^{q^{(k)}_{\textrm{BvK}}}(g_n)=\exp(iq^{(k)}_{\textrm{BvK}}nd)=\exp(iq^{(k)}_{\textrm{BvK}}ld)=%
D_{\Tper}^{q^{(k)}_{\textrm{BvK}}}(h_l). \label{eq:DTd=DTper}
\end{equation} 

In summary, we have introduced a number of groups which are related as follows:
\begin{eqnarray}
\langle\TNd,*\rangle &\subset& \langle\Td,*\rangle, \nonumber \\
\langle \Tper,\boxplus \rangle &=&
\langle \Td,*\rangle / \langle\TNd,*\rangle \cong \langle\Dper,\bigotimes\rangle \cong
\langle\QBvK,\oplus\rangle \subset 
\langle\Qc,\oplus\rangle \cong
\langle\Dd,\bigotimes\rangle.
\end{eqnarray} 

\subsection{``Selection rules'' theorem}
\label{app:SelectionRules}
Consider a function $f(x)=\sum_k\alpha_{k}\Psi_{k}(x)$, which transforms according to an irreducible representation $D$ of some group $G$, i.e. 
\begin{eqnarray}
& f(g^{-1} x) =\sum_{k,j} D(g)_{k,j} \alpha_{j}\Psi_{k}(x),
\,\,\,\, \forall g \,\in\,G. \label{eq:tensor_transform}
\end{eqnarray}
It follows from standard representation theory that
\begin{equation}
D\ne D^0\Rightarrow \int_{\mathcal{M}} dx\,f(x) =0, \label{eq:integral-product}
\end{equation}
where $D^0$ is the identity representation of $G$ and $\mathcal{M}$ is a group invariant manifold. This general theorem \cite{Landau_and_Lifshitz_QM,BirPikus74} is the basis of the quantum mechanic selection rules and follows from the well known orthogonality
relations of group representations \cite{Jansen_and_Boon67,BirPikus74,Landau_and_Lifshitz_QM}.

In the case of $\Tper$ all irreducible representations are one-dimensional, i.e. $f(x)\sim \exp(iq^{(k)}_{\textrm{BvK}}x)$ and $\mathcal{M}\equiv [0,Nd]$. 

In this case one can directly see the validity of Eq.~(\ref{eq:integral-product}):
\begin{eqnarray}
k\ne0\Rightarrow\int_{0}^{Nd}dx\, f_{per}(x) e^{iq^{(k)}_{\textrm{BvK}}x} =\left(  \sum_{l=0}^{N-1}%
\exp(iq^{(k)}_{\textrm{BvK}}ld)\right)  \int_{0}^{d}f_{per}(x) e^{iq^{(k)}_{\textrm{BvK}}x}dx\nonumber\\
=\frac{1-\exp(\pm i2\pi k)}{1-\exp(\pm i2\pi k/N)}\int_{0}^{d}f_{per}(x) e^{iq^{(k)}_{\textrm{BvK}}x}dx=0,
\end{eqnarray}
due to the vanishing numerator. This holds only if $k$ is not a multiple of $N$, which is, of course, fulfilled for $k\in\KBvK$ [cf. Eq. (\ref{eq:q_BvK})].

\subsection{Proof of Equations~(\ref{eq:fundamentalEq1}-\ref{eq:fundamentalEqGeneral})}
\label{app:Proofs}
To prove Eq.~(\ref{eq:fundamentalEq1}) one should take into account that
provided a function $f(x)$ transforms according to an irreducible
representation of a translation group its derivatives
$\frac{d^n}{dx^n}f(x)$ also transform according to the same
representation of this group. This fact follows from the
invariance of the derivative operator with respect to
translations.

The integrand of Eq.~(\ref{eq:fundamentalEq1}) transforms with respect to the infinite translation group $\Td$ as
\begin{equation}
D_{\Td}^0 \otimes \left\lbrace \bigotimes_{r=1}^R D_{\Td}^{q_{r}} \right\rbrace = D_{\Td}^0 \otimes D_{\Td}^{\bigoplus_r q_{r}}= 
D_{\Td}^{q^{(k)}_{\textrm{BvK}}}, \quad k\ne 0 \label{eq:proof1}.
\end{equation} 
From Eq.~(\ref{eq:DTd=DTper}) we know that 
$D_{\Td}^{q^{(k)}_{\textrm{BvK}}}=
D_{\Tper}^{q^{(k)}_{\textrm{BvK}}}$. 
This result together with Eq.~(\ref{eq:integral-product}) completes the proof.

Taking into account that 
$[D_{\Td}^{q_{s}}]^*=D_{\Td}^{-q_{s}}$, Eq.~(\ref{eq:fundamentalEq2}) follows directly from Eq.~(\ref{eq:proof1}).
Eq.~(\ref{eq:fundamentalEqGeneral}) follows also directly from Eq.~(\ref{eq:proof1}) taking additionally into account the translational invariance of the differentiation operator.

It should be noted that the essential step in the proof can be seen as a embedding the problem at hand into another one with appropriate symmetries. The result is then a consequence of the symmetry properties of this second problem.

\end{appendix}

\bibliography{commentsOp,../refs}

\begin{thebibliography}{18}
\bibitem[{fn:({\natexlab{a}})}]{fn:BandEdgeBloch}
\bibinfo{note}{Strictly speaking, at the band edge energies (i.e. when
  $q\in\{0,\pi/d\}$) the general solution \cite{Eastham73} is either
  $\Psi^B_{n,(q)}(x)=\exp(iqx)[\alpha_1 u^{(1)}_{n,q}(x)+\alpha_2
  (xu^{(1)}_{n,q}(x)+u^{(2)}_{n,q}(x))]$ or
  $\Psi^B_{n,(q)}(x)=\exp(iqx)[\alpha_1 u^{(1)}_{n,q}(x)+\alpha_2
  u^{(2)}_{n,q}(x)]$, depending on the symmetry of the unit cell. Here
  $u^{(1)}_{n,q}(x)$ and $u^{(2)}_{n,q}(x)$ are two different periodic
  functions belonging either both to $q=0$ or both to $q=\pi/d$.}
\bibitem[{fn:({\natexlab{b}})}]{fn:SurfaceStates}
\bibinfo{note}{The so-called surface states that arise in systems with CQC lie
  in general energetically outside the band and are characterized by an
  imaginary Bloch wave number. The history of the study of surface states
  together with new insights have been reported recently in \cite{Pereyra05}.}
\end{thebibliography}

\end{document}